\documentclass[preprint,nohyper]{JHEP3} %
\usepackage{epsfig,multicol}
\usepackage{latexsym}

\usepackage{amssymb,amsmath}

\newcommand{\pfrac}[2]{\left( \frac{#1}{#2} \right)}

\newcommand{\unit}[1]{\; \mbox{#1}}

\def\simge{\mathrel{%
   \rlap{\raise 0.511ex \hbox{$>$}}{\lower 0.511ex \hbox{$\sim$}}}}
\def\simle{\mathrel{
   \rlap{\raise 0.511ex \hbox{$<$}}{\lower 0.511ex \hbox{$\sim$}}}}

\newcommand{\dis}[1]{\begin{equation} #1\end{equation}}

\newcommand{\newc}{\newcommand}
\newc\eg{{\it {e.g.}}}  \newc\etal{{\it {et al.}}} \newc\ie{{\it i.e.}}
\newc\etc{{\it {etc}}}  

\newcommand\lsim{\mathrel{\rlap{\lower4pt\hbox{\hskip1pt$\sim$}}
    \raise1pt\hbox{$<$}}}
\newcommand\gsim{\mathrel{\rlap{\lower4pt\hbox{\hskip1pt$\sim$}}
    \raise1pt\hbox{$>$}}}

\newc{\mhalf}{m_{1/2}}      \newc{\mzero}{m_0}

\newc{\tanb}{\tan\beta}
\newc{\azero}{A_0}
\newc{\at}{A_t} \newc{\abot}{A_b} \newc{\atau}{A_\tau} 

\newc{\bmu}{B\mu}           \newc{\sgn}{{\rm sgn}}
\newc{\mone}{M_1}           \newc{\mtwo}{M_2}

\newc{\charone}{\chi_1^\pm} \newc{\mcharone}{m_{\chi_1^\pm}}

\newc{\hl}{h}               \newc{\mhl}{m_{\hl}}
\newc{\hh}{H}               \newc{\mhh}{m_{\hh}}
\newc{\ha}{A}               \newc{\mha}{m_{\ha}}
\newc{\hc}{H^{\pm}}         \newc{\mhc}{m_{\hc}}

\newc{\mw}{m_{W}}      \newc{\mz}{m_{Z}}

\newc{\mgut}{M_{\rm GUT}}

\newc{\mplanck}{M_{\rm P}}      \newc{\mpl}{M_{\rm Pl}}
\newc{\msusy}{M_{\rm SUSY}}      \newc{\ms}{M_{\rm S}}

\newc{\jxf}{J({\xf})}
\newc{\jxfexact}{J_{\rm exact}({\xf})}  \newc{\jxfexp}{J_{\rm exp}({\xf})}
\newc{\VEV}[1]{\langle #1 \rangle}

\newc{\xf}{x_f}
\newc\vrel{v_{\rm rel}}
\newcommand\mchi{m_{\chi}}              

\newc\sell{{\widetilde e}_L}      \newc\msell{m_{\sell}}
\newc\selr{{\widetilde e}_R}      \newc\mselr{m_{\selr}}

\newc\snue{{\widetilde \nu}_e}      \newc\msnue{m_{\snue}}
\newc\snutau{{\widetilde \nu}_\tau}      \newc\msnutau{m_{\snutau}}

\newc\supl{{\widetilde u}_L}      \newc\msupl{m_{\supl}}
\newc\supr{{\widetilde u}_R}      \newc\msupr{m_{\supr}}
\newc\sdl{{\widetilde d}_L}      \newc\msdl{m_{\sdl}}
\newc\sdr{{\widetilde d}_R}      \newc\msdr{m_{\sdr}}

\newcommand{\stauone}{{\tilde \tau}_1}   \newcommand\mstauone{m_{\stauone}}

\newcommand\stau{\tilde{\tau}}
\newcommand\mstau{m_{\stau}}

\newcommand\gluino{\tilde g}
\newcommand\mgluino{m_{\gluino}}

\newc\hpm{H^\pm} \newc\hp{H^+} \newc\hm{H^-} 
\newc\sfermion{\tilde f}  \newc\msfermion{m_{\sfermion}}  

\newc\second{{\rm sec}} 

\newc\alphas{\alpha_s}

\newc\alphaem{\alpha_{em}}

\newcommand\treh{T_{\rm R}}

\newc{\sthw}{\sin\theta_W}              \newc{\cthw}{\cos\theta_W}
\newc{\bino}{\widetilde B}              \newc{\wino}{\widetilde W_3}
\newc{\higgsinob}{{\widetilde H}^0_b}   \newc{\higgsinot}{{\widetilde H}^0_t}

\newc{\abund}{\Omega h^2}
\newc{\abundchi}{\Omega_\chi h^2}
\newc{\abundcdm}{\Omega_{{\rm CDM}} h^2}

\newc{\abundlsp}{\Omega_{\rm LSP}h^2}
\newc{\abundlsptp}{\Omega_{\rm LSP}^{\rm TP}h^2} 
\newc{\abundlspntp}{\Omega_{\rm LSP}^{\rm NTP}h^2}
\newc{\omegam}{\Omega_{{\rm M}}}       \newc{\abundm}{\Omega_{{\rm M}} h^2}
\newc{\omegab}{\Omega_{{\rm b}}}	\newc{\abundb}{\Omega_{{\rm b}} h^2}
\newc{\omegatot}{\Omega_{{\rm TOT}}}

\newc{\omeganlsp}{\Omega_{{\rm NLSP}}}   \newc{\abundnlsp}{\Omega_{\rm NLSP}h^2}
\newc{\ynlsp}{Y_{{\rm NLSP}}}            \newc{\taunlsp}{\tau_{{\rm NLSP}}}
\newc{\nnlsp}{n_{{\rm NLSP}}}            \newc{\mnlsp}{m_{{\rm NLSP}}}
\newc{\nx}{n_{X}}                        \newc{\yx}{Y_{X}}
\newc{\mx}{m_{X}}                        \newc{\taux}{\tau_{X}}

\newc{\rhocrit}{\rho_{crit}}
\newc{\rhochi}{\rho_{\chi}}

\newcommand\neut{\tilde \chi}

\newc{\cachigamma}{C_{a\neut\gamma}}
\newc{\caww}{C_{aWW}}                   
\newc{\cayy}{C_{aYY}}

\newc{\nl}{\cos \theta_{\tilde t}}
\newc{\nr}{\sin \theta_{\tilde t}}
\newcommand\tev{\,\mbox{TeV}}
\newcommand\gev{\,\mbox{GeV}}
\newcommand\mev{\,\mbox{MeV}}
\newcommand\kev{\,\mbox{keV}}

\newc\gbar{{\overline{g}}}

\newc{\ra}{\rightarrow}

\newc{\beq}{\begin{equation}}
\newc{\eeq}{\end{equation}}
\newc{\bea}{\begin{eqnarray}}
\newc{\eea}{\end{eqnarray}}
\newc{\nspin}{n_{\rm spin}}
\newc{\nflavor}{n_{\rm F}}
\newc{\ngamma}{n_\gamma}
\newc{\ychi}{Y_{\chi}}                  \newc{\yeqchi}{Y^{\rm EQ}_{\chi}}

\newcommand\axino{\tilde{a}}

\newc{\naxino}{n_{\axino}}
\newc{\yaxino}{Y_{\axino}}
\newc{\yeqaxino}{Y^{\rm EQ}_{\axino}}
\newc{\ythaxino}{Y^{\rm TP}_{\axino}}
\newc{\ynthaxino}{Y^{\rm NTP}_{\axino}}

\newcommand\gravitino{\widetilde{G}}    
\newcommand\mgravitino{m_{\gravitino}}

\newcommand\abundg{\Omega_{\gravitino}h^2}
\newcommand\abundgntp{\Omega^{\rm NTP}_{\gravitino}h^2}     

\newcommand\abundgtp{\Omega^{\rm TP}_{\gravitino}h^2}       

\newc{\ngravitino}{n_{\gravitino}}
\newc{\ygravitino}{Y_{\gravitino}}
\newc{\yeqgravitino}{Y^{\rm EQ}_{\gravitino}}
\newc{\ythgravitino}{Y^{\rm TP}_{\gravitino}}
\newc{\ynthgravitino}{Y^{\rm NTP}_{\gravitino}}

\newc{\yascat}{Y^{\rm scat}_{i,j}}      \newc{\yadec}{Y^{\rm dec}_{i}}
\newc{\gstar}{g_\ast}           \newc{\gsstar}{g_{s\ast}}

       \def\pslash{\not{\hbox{\kern-2.3pt $p$}}}
       \def\kslash{\not{\hbox{\kern-2.3pt $k$}}}
       \def\qslash{\not{\hbox{\kern-2.3pt $q$}}}
       \def\ddslash{\not{\hbox{\kern-2.3pt $d$}}}
       \def\prtslash{\not{\hbox{\kern-2.3pt $\partial$}}}

\newcommand\jcap[3] %   {\@spires{PHRVA%2CD#1%2C#3}
		{{\it J.\ Cosmol.\ Astrop.\ Phys.\ }{\bf #1} (#2) #3}
\title{A Re-analysis of Gravitino Dark Matter in the Constrained MSSM}
\author{Sean Bailly\\
Laboratoire de Physique Th\'eorique et Astroparticules,
CNRS UMR 5825,\\
Universit\'e Montpellier II, F-34095 Montpellier Cedex 5, France\\
E-mail: \email{Sean.BAILLY@lpta.univ-montp2.fr}
} 
\author{Ki-Young Choi\\
        Departamento de F\'{\i}sica Te\'{o}rica C-XI,
        Universidad Aut\'{o}noma de Madrid, Cantoblanco,
        28049 Madrid, Spain\\   
        Instituto de F\'{\i}sica Te\'{o}rica UAM/CSIC,
        Universidad Aut\'{o}noma de Madrid, Cantoblanco,
        28049 Madrid, Spain\\
        E-mail: \email{kiyoung.choi@uam.es}}
\author{Karsten Jedamzik\\
Laboratoire de Physique Th\'eorique et Astroparticules,
CNRS UMR 5825,\\
Universit\'e Montpellier II, F-34095 Montpellier Cedex 5, France\\
E-mail: \email{jedamzik@LPM.univ-montp2.fr}
}
\author{Leszek Roszkowski\footnote{On leave of absence from The
    Andrzej Soltan Institute for Nuclear Studies, Warsaw, Poland.}\\
Department of Physics and Astronomy,
University of Sheffield, Sheffield, S3 7RH, UK\\
E-mail: \email{L.Roszkowski@sheffield.ac.uk}
}

\abstract{We re-consider the gravitino as dark matter in the
framework of the Constrained MSSM. We include several recently
suggested improvements on: (i) the thermal production of gravitinos,
(ii) the calculation of the hadronic spectrum from
NLSP decay and (iii) the BBN calculation including stau bound-state
effects. In most cases we find an upper bound on the reheating
temperature $\treh\lsim {\rm ~a~few}\times 10^7\gev$ from
over-production of $^6{\! \rm Li}$ from bound state effects.  We also
find an upper limit on the stau lifetime of $3\times10^4 \sec$, which
is nearly an order of magnitude larger than the
simple limit $5\times 10^3 \sec$ often used 
to avoid the effect of bound-state catalysis.  The bound on $\treh$ is relaxed to
$\lsim {\rm ~a~few}\times 10^8\gev$ when we use a more conservative
bound on ${^6}{\! \rm Li}/{^7}{\!\rm Li}$, in which case a new region
at small stau mass at $\sim 100\gev$ and much longer lifetimes opens
up. Such a low stau mass region can be easily tested at the LHC. }
\preprint{FTUAM 09/3, IFT-UAM/CSIC-09-15}

\begin{document}

%=====================
\section{Introduction}\label{sect:intro}
%\paragraph{Introduction.}
%=====================

Local supersymmetry (SUSY), or supergravity, models predict the
existence of a massive spin-3/2 particle, the gravitino $\gravitino$,
whose mass in general depends on a supersymmetry breaking mechanism.
Assuming standard big bang cosmology, it was shown early on that
cosmological constraints require that the mass of gravitino
$\mgravitino$ be much less than $1 \kev$, or else heavier than some
$10 \tev$~\cite{pp82,weinberg-grav82}.  While a primordial
gravitino population can efficiently be diluted by
inflation~\cite{Ellis:1982yb}, subsequently the Universe can be
repopulated with gravitinos via thermal production (TP) processes
involving scatterings of Standard Model (SM) and SUSY particles in the hot plasma, with
the number density proportional to the reheating temperature,
$\treh$. If, assuming $R$-parity, the gravitino is the lightest supersymmetric particle
(LSP) and stable, one can also generate gravitinos via non-thermal
production (NTP) process of freeze-out and decays of the next-to-lightest
superpartner (NLSP). Because of the gravitino's exceedingly weak
couplings to ordinary matter, the latter process usually takes place
during or after the period of Big Bang Nucleosynthesis (BBN) and
involves releasing substantial amounts of electromagneticaly and/or 
hadronically interacting particles 
(the importance of the latter shown to be important
in~\cite{Jedamzik:2004er,kkm04,Jedamzik:2006xz}), which could wreck havoc to successful
predictions of Big Bang Nucleosynthesis (BBN). In order to avoid this,
and assuming gravitinos to
be a dominant component of dark matter (DM), one imposes an
upper bound on the reheating temperature of
$\treh<10^{6-8}\gev$~\cite{nos83,khlopov+linde84,ekn84,ens84,jss85,km94}
(for recent updates see,
\eg,~\cite{cefo02,kkm04,rrc04,Kohri:2005wn,Cerdeno:2005eu,Jedamzik:2006xz,Kawasaki:2008qe}).
 
Considering NTP processes only, constraints on the parameter space of
Minimal Supersymmetric Standard Model (MSSM) were derived
in~\cite{feng03-prl,feng03,fst04}, basically eliminating the lightest
neutralino as the NLSP (unless
$\mgravitino\lsim1\gev$~\cite{Cerdeno:2005eu}).  In~\cite{rrc04} some
of us considered a combined impact of both thermal and nonthermal
production mechanisms in the more predictive framework of the
Constrained MSSM (CMSSM)~\cite{kkrw94} and derived both the maximum
allowed reheating temperature and the viable regions of parameters of
the model. An improved analysis using full systematic BBN calculation
was next conducted in~\cite{Cerdeno:2005eu}. Both showed, in
particular, that gravitino dark matter abundance from NTP
processes alone~\cite{feng03-prl,feng03,eoss03-grav,fst04} can only
agree with observations in regions of SUSY parameters where a gluino
mass is very heavy, in the multi-$\tev$ range, substantially above an
approximate value derived in~\cite{fiy03}. In other words, in order to
have gravitino DM consistent with superpartner masses at or below the
$\tev$ scale, a substantial TP contribution to the total abundance is
necessary. (In~\cite{Cerdeno:2005eu} it was also shown that the
stau NLSP region consistent with the cosmological abundance of DM in
gravitinos corresponds to a false (local) vacuum of the CMSSM.)
Possible solutions to cosmic lithium problems were also investigated
in~\cite{Jedamzik:2005dh}.  The papers~\cite{rrc04,Cerdeno:2005eu}
were next followed by similar detailed
analyzes~\cite{Steffen:2006hw,Pradler:2006qh}.

More recently it was pointed out~\cite{Pospelov:2006sc} that the
existence of unstable (with lifetime longer than $10^3$ seconds), 
negatively charged electro-weak scale particles alter the predictions
for lithium and other light element abundances via the formation of metastable
bound states with nuclei during
BBN~\cite{Pospelov:2006sc,Kaplinghat:2006qr,Kohri:2006cn,
Hamaguchi:2007mp,Jedamzik:2007qk}.  These bound state effects
were incorporated in the gravitino dark matter with charged particle
NLSP~\cite{Cyburt:2006uv,Pradler:2006hh,Pradler:2007is,Kersten:2007ab}.

In light of these recent developments, we present here an updated
analysis of the gravitino DM in the Constrained MSSM. Relative to the
previous works~\cite{rrc04,Cerdeno:2005eu}, we make improvements 
in the following aspects: (i) the thermal production of gravitinos
including the decay allowed by thermal masses~\cite{Rychkov:2007uq} as
well as scattering processes; (ii) the calculation of hadronic and
electromagnetic spectrum from NLSP decay including correct
implementation of left-right stau mixing; (iii) the updated BBN
calculation including the bound-state effects of charged massive
particles.

As previously in~\cite{rrc04,Cerdeno:2005eu}, we will treat
$\mgravitino$ as a free parameter and will not address the question of
an underlying supergravity model and SUSY breaking mechanism.  We will
allow $\mgravitino$ to vary over a wide range of values from ${\cal
O}(\tev)$ down to the ${\cal O}(\gev)$ scale (as most natural in the
CMSSM with gravity-mediated SUSY breaking), for which gravitinos (at
least those produced in thermal production) would remain mostly cold DM
relic, but will explore at some level also lighter gravitinos, down to
$100\mev$. Less massive gravitinos could remain DM in some models of
gauge mediated supersymmetry
breaking~\cite{Fujii:2002yx,Fujii:2002fv,Lemoine:2005hu,Jedamzik:2005ir}.
 
The paper is organized as follows. 
In section~\ref{sect:improvements} we describe the improvements made
in the present analysis compared to the previous ones.
In section~\ref{sect:results} we summarize the experimental and cosmological 
constraints, show our numerical results and derive an upper bound on $\treh$.
We summarize our conclusions in section~\ref{sect:summary}.

\newpage
%=====================
\section{Improvements in the analysis}
\label{sect:improvements}
%=====================
In this section we list  the improvements made in this paper.

\subsection{Gravitino production}

We will consider gravitinos as DM relics which were produced
predominantly via the TP and NTP mechanisms stated above.\footnote{In addition,
there could be other possible ways of populating the Universe
with stable gravitinos, \eg\ via inflaton decay or during
preheating~\cite{grt99,kklp99,Kawasaki:2006gs,Endo:2007sz}, 
or from decays of moduli fields~\cite{kyy04}. 
In some of these cases the gravitino production
is independent of the reheating temperature and its abundance may give the
measured dark matter abundance with no ensuing limit on $\treh$. In
general, such processes are, however, much more model dependent and
not necessarily efficient~\cite{nps01}, and will not be considered
here.}

Since the original computation of TP rates in~\cite{ekn84} there have
been a number of updates and
improvements~\cite{mmy93,Kawasaki:1994af,bbp98,bbb00,Pradler:2006qh,Rychkov:2007uq}. 
In particular, a gauge invariant computation was performed
in~\cite{bbb00} (with an extension to the three SM gauge
groups done in~\cite{Pradler:2006qh}), and a different technique to compute
gravitino production from decays allowed by thermal masses and the
effect of the top Yukawa coupling were applied
in~\cite{Rychkov:2007uq}.
In this update, we adopt the result of~\cite{Rychkov:2007uq},
including the three gauge groups, which leads to  about a factor of two
enhancements compared to the previous calculation~\cite{bbp98,bbb00}.
For the renormalized gauge couplings and gaugino masses  
at an energy scale $\treh$, we used a one-loop evolution by 
the renormalization group equation in the MSSM from GUT scale
assuming the gauge coupling and gaugino mass unification,
%%%
\dis{
M_i(T)=\left(\frac{g_i(T)}{g_{GUT}}\right)^2 m_{1/2}.
\label{gauginooneloop}
} 
%%%
The thermal production will therefore depend on the common gaugino
mass $\mhalf$ and the other parameters of the CMSSM: the common scalar
mass $\mzero$, $\tanb$ and the trilinear soft scalar coupling
$\azero$, as well as on the reheating temperature $\treh$. 
In our
analysis we use an expression for $\abundgtp$ as computed in
ref.~\cite{Rychkov:2007uq}
%
%%%
\beq
\abundgtp=
0.167\left(\frac{\mgravitino}{100\gev}\right)\left(\frac{\treh}{10^{10}\gev}\right)\left(\frac{\gamma 
  \, (\treh)}{\treh^6/\mplanck^2}\right), 
\label{abundgtp:eq}
\eeq 
%%%
where the gravitino production rate $\gamma$ is the sum of three contributions
from decay, subtracted scattering and top Yukawa coupling,
%%%
\dis{
\gamma = \gamma_D + \gamma_S^{sub} + \gamma_{top},
}
%%%
and the details are given in ref.~\cite{Rychkov:2007uq}. 

Regarding the non-thermal production of gravitinos, we proceed in the usual way.
Since all the NLSP particles decay after freeze-out, the gravitino
relic abundance from NTP, $\abundgntp$, is related to $\Omega_{\rm NLSP} h^2$ --
the relic abundance that the NLSP would have had if it had remained
stable -- via a simple mass ratio
\beq
\abundgntp=\frac{\mgravitino}{\mnlsp}\Omega_{\rm NLSP} h^2.
\label{abundgntp:eq}
\eeq Note that $\abundgntp$ grows with the mass of the gravitino
$\mgravitino$.  When we calculate the relic number density of staus, we
treat correctly the mixing of left and right stau states using
a supersymmetric particle spectrum calculator
SuSpect~\cite{Djouadi:2002ze} and employ micrOMEGAs~\cite{Belanger:2006is} 
to calculate a relic density of NLSPs, which might affect the
annihilation cross section when the maximal mixing happens at large
$\tanb$, as noticed by~\cite{Kersten:2007ab}.

The total abundance of the LSPs is the sum of both thermal and non-thermal 
production contributions
\dis{
\abundg = \abundgtp + \abundgntp.
\label{Omega_tot}
}
Since it is natural to expect that the LSP makes up most of DM in the
Universe, we can re-write the above as~\cite{Choi:2007rh}
\dis{
\abundg =\abundgtp\left(\treh,\mgravitino,\mgluino,\mnlsp,\ldots \right) +
\frac{\mgravitino}{\mnlsp} \abundnlsp = \Omega_{\rm DM}\simeq 0.1.
\label{eq:oh2relation}
}
%

%=====================
\subsection{Hadronic and electromagnetic spectrum from NLSP decay}
%=====================

In order to calculate light element abundances during BBN, including
the impact of massive particle decays, in our case the NLSP, we need to
know the detailed decay products and their spectrum in addition to NLSP
lifetime and relic number density.

The lifetime of the stau or neutralino NLSP decaying to the gravitino 
is dominated by  two-body decays,
\begin{eqnarray}
\Gamma_{\rm tot}^{\stau_1} &\simeq& \Gamma(\stau_1\rightarrow \tau \gravitino), \\
\Gamma_{\rm tot}^{\chi} &\simeq& \Gamma(\chi \rightarrow \gravitino
\gamma) + \Gamma(\chi \rightarrow \gravitino Z) + \Gamma(\chi
\rightarrow \gravitino h), 
\end{eqnarray}
where, for the stau NLSP,
\begin{equation}
\Gamma(\stau_1\rightarrow \tau \gravitino) =
\frac{1}{48\pi}\frac{m_{\stau}^5}{M_{\rm Pl}^2 \mgravitino^2}
\left( 1-\frac{\mgravitino^2}{m_{\stau}^2}\right)^4, 
\label{eq:lifegrav}
\end{equation}
and for the neutralino NLSP the dominant decay to photon and gravitino
is given by~\cite{fst04}
\begin{eqnarray}
\Gamma(\chi \rightarrow \gamma\gravitino) &=& \frac{|N_{11}\cos
  \theta_W + N_{12}\sin \theta_W|^2}{48\pi}
\frac{\mchi^5}{M_{\rm Pl}^2\mgravitino^2} \left(
1-\frac{\mgravitino^2}{\mchi^2}\right)^3
\left(1+3\frac{\mgravitino^2}{m_{\chi}^2} \right). 
\end{eqnarray}
These decay widths depend mostly on the LSP and NLSP masses, but
also on $N_{ij}$ , the neutralino mass mixing matrix.  Note that
$M_{\rm Pl}$ denotes the reduced Planck mass. The
lifetime can be sufficiently long $\tau \sim 10^2 - 10^5 \unit{s}$ to
be interesting for BBN and the lithium anomalies.

The two-body decays also dominate the electromagnetic cascade with the
decay of the tau-lepton  (producing either directly leptons or mesons
that decay electromagnetically before interacting strongly with the
light elements). The electromagnetic branching ratio and energy
injected in the cascade for the stau are given by:
\begin{eqnarray}
B_{\rm em}^{\stau} = \frac{\Gamma(\stau_1\rightarrow
  \tau \gravitino)}{\Gamma_{\rm tot}} \simeq 1,\qquad\qquad\qquad
E_{\rm em}^{\stau} \simeq
\frac{1}{2}\pfrac{m_{\stau_1}^2-\mgravitino^2}{2m_{\stau}}. 
\end{eqnarray}
The approximate fraction $1/2$ in the total tau energy is a reflection of the fact
that some energy is lost into non-interacting neutrinos. 
Similar results are obtained for the neutralino:
\begin{eqnarray}
B_{\rm em}^{\chi} \simeq 1,\qquad\qquad\qquad
E_{\rm em}^{\chi} = \pfrac{m_{\chi}^2-\mgravitino^2}{2m_{\chi}}.
\end{eqnarray}
Hadronic energy contributions come from 3-body, or from 4-body,
NLSP decays:
\begin{eqnarray}
\stau_1 &\rightarrow& \tau \gravitino Z/\gamma \rightarrow \tau
\gravitino q\bar{q}, \label{eq:decayZqqbar}\\ 
\stau_1 &\rightarrow& \nu_\tau \gravitino W \rightarrow \nu_\tau
\gravitino q\bar{q}, \label{eq:decayWqqbar}\\ 
\stau_1 &\rightarrow& \tau \gravitino h \rightarrow \tau
\gravitino q\bar{q}. \label{eq:decayhqqbar} 
\end{eqnarray}
In our previous analyzes~\cite{rrc04,Cerdeno:2005eu} we approximated
these processes by a simple formula, and the
process~(\ref{eq:decayWqqbar}) was set to zero, since we assumed 
the stau  to be a pure right-handed state. Here  we incorporate the mixing and use the
programs, CalcHEP \cite{Pukhov:2004ca} (an automatic matrix element
generator) and PYTHIA \cite{Sjostrand:2006za} (a Monte-Carlo
high-energy-physics event generator) to produce the spectrum of the 
decay products properly.  With this update eq.~(\ref{eq:decayWqqbar}) is not
zero any more due to the left-handed component enabling this
process.

Following~\cite{Steffen:2006hw} we use the $q\bar{q}$ invariant mass 
$m_{q\bar{q}}$ to calculate the decay width in eq.~(\ref{eq:decayZqqbar}).
For completeness virtual Higgs exchanges were also calculated. The
decay width is given with a cut on the $q\bar{q}$ invariant mass
$m_{q\bar{q}}^{\rm cut}=2\unit{GeV}$ below which no nucleon can be
produced,
\begin{equation}
\Gamma(\stau \rightarrow \tau \gravitino q
\bar{q};m_{q\bar{q}}^{\rm cut}) = \int_{m_{q\bar{q}}^{\rm
    cut}}^{m_{\stau}-\mgravitino-m_{\tau}}
dm_{q\bar{q}}\frac{d\Gamma(\stau \rightarrow \tau \gravitino q
  \bar{q})}{dm_{q\bar{q}}}. 
\end{equation}
The branching ratio is given by
\begin{equation}
B_h(\stau \rightarrow \tau \gravitino q
\bar{q};m_{q\bar{q}}^{\rm cut})=\frac{\Gamma(\stau \rightarrow
  \tau \gravitino q \bar{q};m_{q\bar{q}}^{\rm cut})}{\Gamma_{\rm
    tot}}. 
\end{equation}
%We also follow the proposition in Ref.~\cite{Steffen:2006hw}
%to use the average of the invariant mass 
%$\langle m_{q\bar{q}}\rangle$ as an estimator of the "effective hadronic
%energy" for the BBN calculations. 
%However, as discussed in more detail in Ref.~\cite{bailly:yy},
%even this presents an approximation. Fully accurate calculations are
%feasible but very time-consuming.  
%
The hadronic decay is dominated by an exchange of $Z/\gamma$. The $W$
processes are suppressed by one or two orders of magnitude compared to
the $Z/\gamma$ because the left-handed component is comparatively
small in the CMSSM.  The Higgs contribution is even more negligible
due to the Higgs mass in the propagator. This last contribution can be
enhanced at large $\tanb$ which increases the Higgs boson couplings
but it remains below the $W$ contribution. For the neutralino NLSP,
the hadronic cascades come from 3-body decay $\chi \rightarrow
\gravitino q\bar{q}$:
\begin{eqnarray}
\chi &\rightarrow& \gravitino Z/\gamma \rightarrow \gravitino q\bar{q}, \\
\chi &\rightarrow& \gravitino h \rightarrow \gravitino q\bar{q},
\end{eqnarray}
and the decay width is calculated similarly as for the stau
\begin{equation}
\Gamma(\chi \rightarrow \gravitino q
\bar{q};m_{q\bar{q}}^{\rm cut}) = \int_{m_{q\bar{q}}^{\rm
    cut}}^{m_{\chi}-\mgravitino} dm_{q\bar{q}}\frac{d\Gamma(\chi
  \rightarrow \gravitino q \bar{q})}{dm_{q\bar{q}}}. 
\end{equation}

\subsection{BBN calculation}

As in our previous work~\cite{Cerdeno:2005eu}, for each point in
supersymmetric parameter space we perform a complete BBN calculation
of light element abundances with NLSP decay-induced cascades.  
These calculations include all thermal
nuclear reactions as well as all (nonthermal) interactions important
for the developments of the electromagnetic and hadronic cascades.
Details about the effects of the around $100$ nonthermal interactions
required to attain an estimated BBN yield accuracy of $30\%$ may be
found in~\cite{Jedamzik:2006xz}. The calculations are performed at
$\Omega_b h^2 = 0.02273$, as inferred by WMAP~\cite{Dunkley:2008ie}.

The improvements with respect to ref.~\cite{Cerdeno:2005eu} are
twofold. Firstly, the current BBN calculations take full and accurate
account of catalytic
effects~\cite{Pospelov:2006sc,Cyburt:2006uv,Kaplinghat:2006qr,Kohri:2006cn,
Hamaguchi:2007mp,Bird:2007ge,Jittoh:2007fr,Jedamzik:2007cp,
Jedamzik:2007qk,Pospelov:2007js,Kamimura:2008fx}. Secondly, the
energies of injected nucleons are treated in a much improved way. It
has been recently shown that the formation of bound states between
electrically charged NLSPs (here principally the stau $\stau$)
and nuclei towards the end of BBN lead to drastic changes of some
thermal nuclear
rates~\cite{Pospelov:2006sc,Hamaguchi:2007mp,Pospelov:2008ta,
Kamimura:2008fx}.  Various modifications have been proposed, but the
by far most important one is the replacement of the weak quadrupole
transition ${\rm D} +{}^4{\rm He}\to {}^6{\rm Li}+\gamma$ with the
catalytic process ${\rm D} +({}^4{\rm He}-\stau)\to {}^6{\rm
Li}+\tau$, with a rate $\sim 10^7$ times larger than the former. Since
both $({}^4{\rm He}-\stau)$ bound states as well as $^6$Li
first survive photo-disintegration/nuclear destruction at $\simge
10^4\sec$, $^6$Li overproduction rules out most of the stau-NLSP
parameter space with stau decay time $\tau\simge
5\times 10^3\sec$~\cite{Pospelov:2006sc,Cyburt:2006uv,Pradler:2006hh}.
The exception here are very small freeze-out NLSP stau-to-entropy
ratios $Y=n /s\simle 10^{-15}-10^{-14}$
~\cite{Jedamzik:2007qk} as recently proposed to occur at large
$\tanb$~\cite{Ratz:2008qh}.  Other modifications to BBN due to 
the bound state induced catalytic effects are the possibility of
appreciable $^9$Be synthesis~\cite{Pospelov:2007js} which has been
used to derive limits on abundances of stau
NLSPs~\cite{Pospelov:2008ta}.  Though we fully calculate $^9$Be/H
ratios, we refrain from using them for limits, as the catalytic rates
seem currently very uncertain~\cite{Kamimura:2008fx}.  Finally, late-time
destruction of $^6$Li and $^7$Li through reactions on
($p-\stau$) bound states, as proposed in
ref.~\cite{Jedamzik:2007cp}, has proven unimportant when the newly
determined catalytic rates of ref.~\cite{Kamimura:2008fx} are
employed.

In ref.~\cite{Cerdeno:2005eu} a partition of energy into hadronic three-
and four- body decays, such as $\stau\to\gravitino\tau Z$ and
$\stau\to\gravitino\tau q\bar{q}$ was approximated by
$E_{Z}\approx E_{q\bar{q}}\approx (m_{\stau}-m_{\gravitino})/3$.
In ref.~\cite{Steffen:2006hw} both processes were treated in detail
and it was found that the invariant mass-squared  $m_{q\bar{q}}^2 =
(P_q+P_{\bar{q}})^2$ of produced $q\bar{q}$ flux tubes is
peaked around the $Z$-mass squared $m_Z^2$. 
  It was argued that the average of
$\langle m_{q\bar{q}}\rangle$ should be used as an ``effective hadronic
energy'' for BBN calculations rather than
$E_{q\bar{q}}\approx(m_{\stau}-m_{\gravitino})/3$.  In
particular for heavy $m_{\stau}\sim 1\,$TeV this would make a
factor $\sim 3$ difference in hadronic effects.  As described in more
detail in ref.~\cite{Bailly:2008yy}, the situation is actually more
complicated.  In fact $\langle m_{q\bar{q}}\rangle$ is {\it not} a
good estimate of the energy of the $q\bar{q}$ flux-tube in the cosmic
rest-frame. Nevertheless, for independent reasons, at early times,
$\tau\simle 300\sec$ taking $\langle m_{q\bar{q}}\rangle$ is indeed a
better approximation, which has been adopted in the present analysis.  A
completely accurate determination would require knowledge of the
detailed primary nucleon energy spectrum due to the NLSP decays. This,
is however, numerically currently not feasible for the large number of
models to be calculated.  Some detailed calculations have been
performed in ref.~\cite{Bailly:2008yy} indicating a $\sim 30\%$
uncertainty in the results.

%=====================
\section{Constraints and results}
%=====================
\label{sect:results}

Unless otherwise stated, we will follow the analysis and notation of
our previous papers~\cite{rrc04,Cerdeno:2005eu} to which we refer
the reader for more details.  Here we only summarize the main results
and differences.

Mass spectra of the CMSSM are determined in terms of the usual four
free parameters mentioned above: $\tanb$, $\mhalf$, $\mzero$, and
$\azero$, as well as on $\sgn(\mu)$ -- the sign of the supersymmetric
Higgs/higgsino mass parameter $\mu$.  The parameter $\mu^2$ is derived
from the condition of the electroweak symmetry breaking and we take
$\mu>0$.

We calculate the thermal contribution $\abundgtp$ following the result
of~\cite{Rychkov:2007uq}.  In evaluating the non-thermal part
$\abundgntp$, we first compute the number density of the NLSP after
freeze-out (the neutralino or the stau) with high accuracy by
numerically solving the Boltzmann equation including all (dominant and
subdominant) NLSP pair annihilation and coannihilation channels. For a
given value of $\mgravitino$, we then compute $\abundgntp$ via
eq.~(\ref{abundgntp:eq}).

After freeze-out from the thermal plasma at $t\sim10^{-12}\sec$, the NLSPs
decay into gravitinos at late times which strongly depend on the NLSP
composition and mass, on $\mgravitino$ and on the final states of the
NLSP decay~\cite{eoss03-grav,fst04}. 
The exact value of the NLSP lifetime in the
CMSSM further depends on a possible relation between $\mgravitino$ and
$\mhalf$ and/or $\mzero$ but in the parameter space allowed by other
constraints it can vary from $\gsim10^8\sec$ at smaller $\mnlsp$ down
to $10^2\sec$, or even less, for $\mhalf$ and/or $\mzero$ in the
$\tev$ range or for small gravitino mass $\mgravitino$.

When the NLSPs decay, the high energy electromagnetic and hadronic
particles are produced and may interact with background light nuclei
and change the abundances.  Using the output from low energy spectrum
and interactions, we calculate the spectrum of decay products of NLSP
and implement into BBN calculation to predict the primordial light
element abundances. 
Then we require that the light element abundances after decay of NLSP are
within the bounds of those inferred from the observational data.

%%%%%%%%%%%%%%%%%%%%%%%%%%%%%%%%%%%%%%%%%%%%%%%%%%%
\subsection{Collider and cosmological constraints}
%%%%%%%%%%%%%%%%%%%%%%%%%%%%%%%%%%%%%%%%%%%%%%%%%%%
We apply the same experimental bounds as
in~\cite{rrc04,Cerdeno:2005eu}: (i) the lightest chargino mass
$\mcharone >104\gev$; (ii) the lightest Higgs mass $m_h>114.4\gev$;
(iii) $BR(B\rightarrow X_s\gamma) = (3.55 \pm 0.68)\times10^{-4}$ and
(iv) the stau mass bound $\mstauone>87\gev$.  In this
analysis we also use the top quark mass $m_t=172.7\gev$~\cite{top172}.
However slight change of these experimental limit does not modify the
final results, since the BBN constraints (especially from $^6{\rm Li}/^7{\rm Li}$)
are more severe.

For the observational constraints on the primordial 
light element abundances, we use~\cite{Jedamzik:2006xz}
\begin{center}
\begin{tabular}{r c l}
$1.2\times10^{-5}<$ & $\rm D/H$ & $< 5.3\times10^{-5}$ 
\\ 
& $Y_p$ & $< 0.258$ 
\\
$8.5\times10^{-11}<$ & ${^7}{\!\rm  Li}/H$ & 
\\
& ${^3}{\! \rm He}/{\rm D}$ & $< 1.52$ 
\\
& ${^6}{\! \rm Li}/{^7}{\! \rm Li}$ & $< 0.1 \,(0.66)$.
\end{tabular}
% \caption{ }
\label{lightelements:table} 
\end{center}
For ${^6}{\! \rm Li}/{^7}{\!\rm Li}$ we will also use the conservative
upper limit which is given in the bracket. It allows for the possibility
of stellar ${^6}{\!\rm Li}$ (and ${^7}{\!\rm Li}$) depletion. 
Since ${^6}{\!\rm Li}$ is more fragile than ${^7}{\!\rm Li}$,
post-BBN lithium processing may conceivably reduce the 
${^6}{\!\rm Li}/{^7}{\!\rm Li}$ ratio.
The reader is referred to
Ref.~\cite{Jedamzik:2006xz} for a more detailed discussion of the adopted
limits.  
In the first two figures
below, the regions excluded by BBN constraints will be shaded violet
and marked ``BBN''.  We also show the BBN constraints with the
conservative bound on ${^6}{\! \rm Li}/{^7}{\!\rm Li}$ with blue
dashed lines.

\begin{figure}[!t]
  \begin{center}
  \begin{tabular}{c c}
   \includegraphics[width=0.5\textwidth]{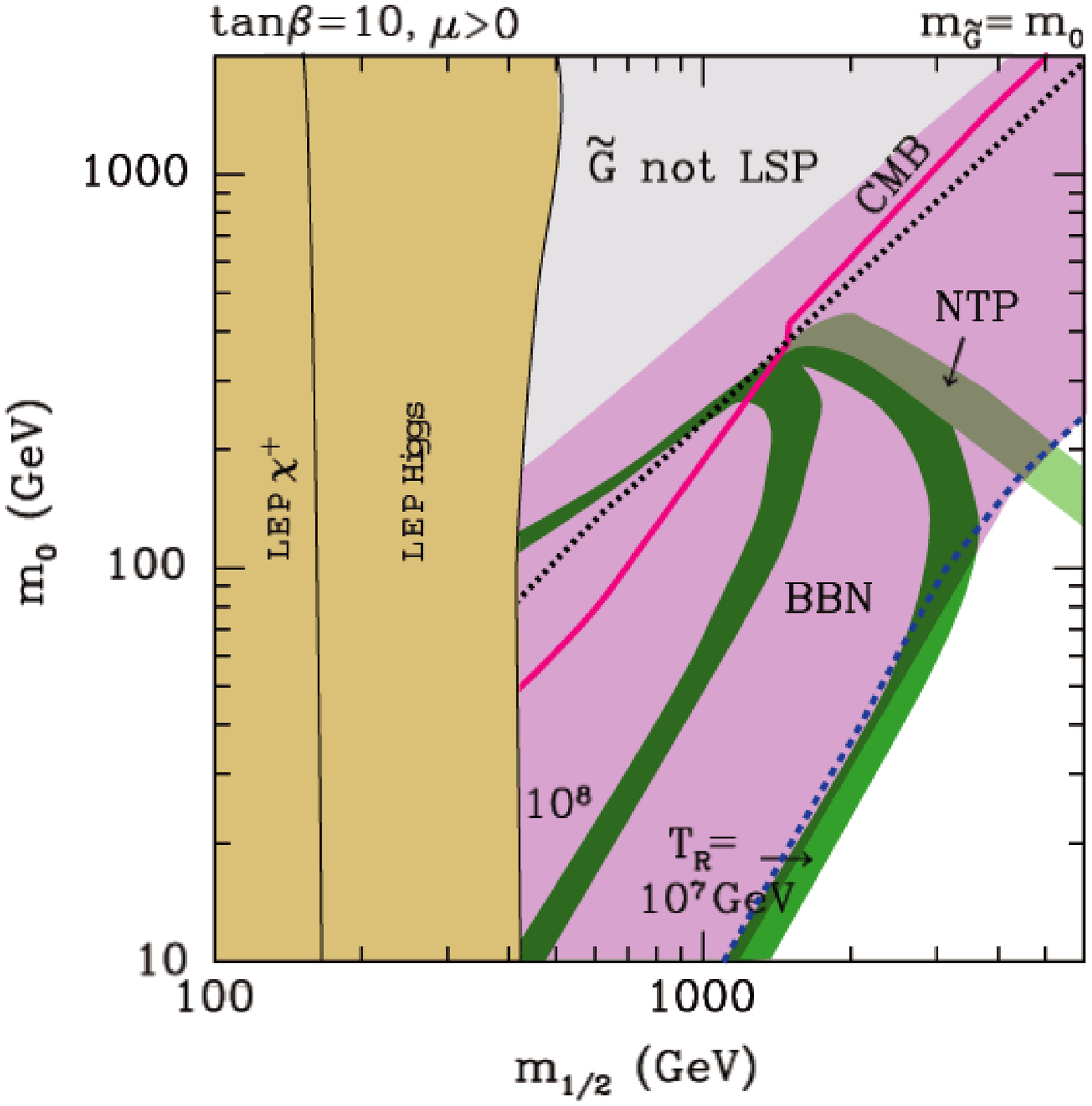}
    & 
  \includegraphics[width=0.5\textwidth]{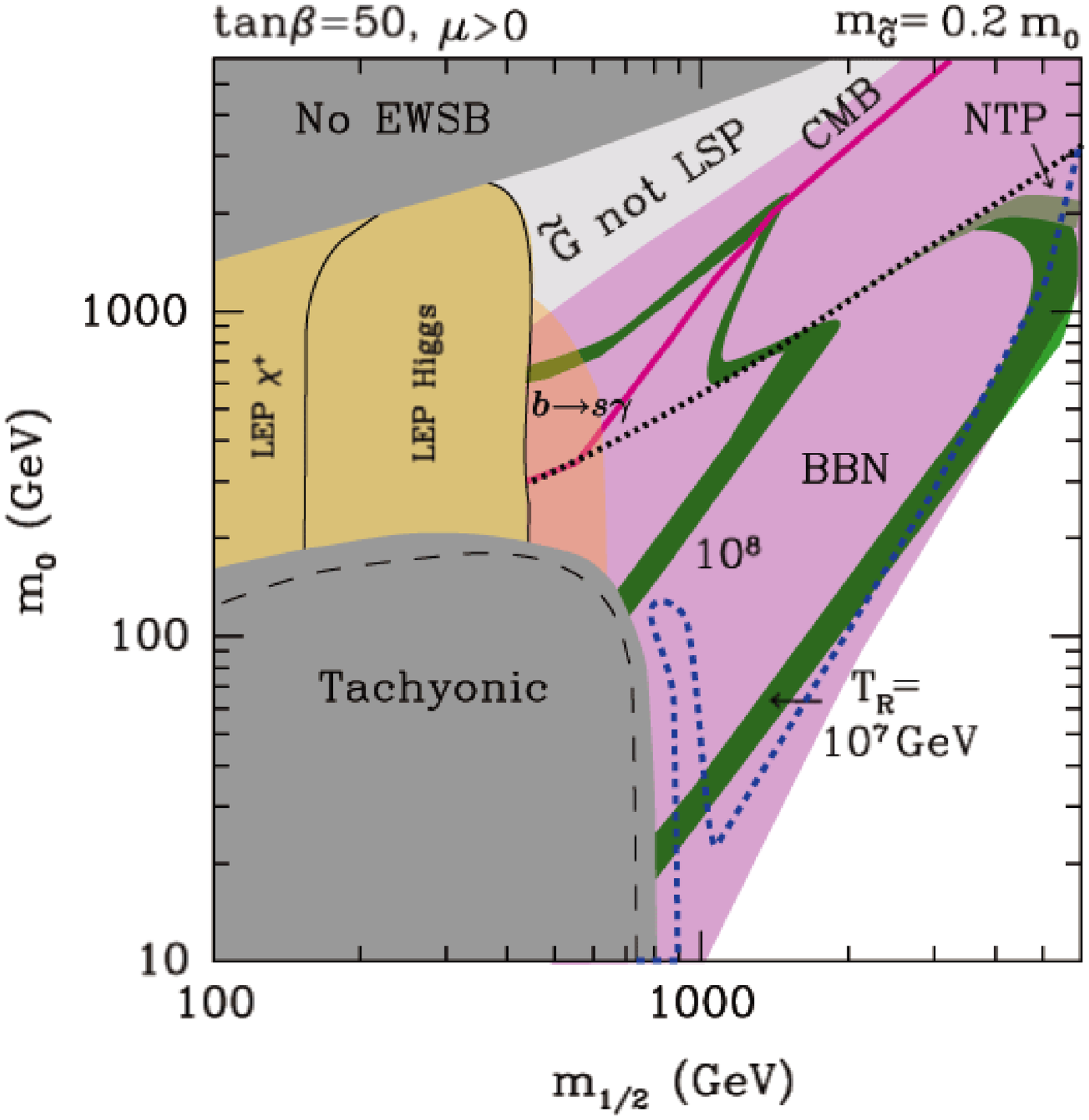}
  \end{tabular}
  \end{center}
  \caption{ \small The plane ($\mhalf,\mzero$) for $\tanb=10$,
$\mgravitino=\mzero$ (left window) and $\tanb=50$,
$\mgravitino=0.2\mzero$ (right window) and for $\azero=0$, $\mu>0$. 
The light brown regions labeled ``LEP $\chi^+$'' and ``LEP Higgs'' are
excluded by unsuccessful chargino and Higgs searches at LEP,
respectively. In the right window the darker brown region labeled
``$b\to s\gamma$'' is excluded assuming minimal flavor violation. The
dark grey region below the dashed line is labeled ``Tachyonic''
because of some sfermion masses becoming tachyonic, and is also
excluded. In the rest of the grey region (above the dashed line) the
stau mass bound $\mstauone>87\gev$ is violated. In the region ``No
EWSB'' the conditions of EWSB are not satisfied.  The dotted diagonal line marks
the boundary between the neutralino ($\chi$) and the stau ($\stau$) NLSP.  The
regions excluded by the various BBN constraints are denoted in violet.
A solid magenta curve labeled ``CMB'' delineates the region (on the
side of label) which is inconsistent with the CMB spectrum.  In both windows,
the dark green bands labeled ``$\treh=10^7 \gev$'' and ``$10^8$''
correspond to the total relic abundance of the gravitino (from the sum of thermal
and non-thermal production), for a given (denoted) reheating
temperature, lying in
the favored range. In the light green regions (marked ``NTP'')
the same is the case for the relic abundance from NTP process alone.
The blue dashed line denotes the relaxed boundary of the BBN constrains when we use
 the conservative  limit ${^6}{\! \rm Li}/{^7}{\!\rm Li}<0.66$.   }
  \label{fig:bbn}
\end{figure}
\begin{figure}[!t]
  \begin{center}
  \begin{tabular}{c c}
   \includegraphics[width=0.5\textwidth]{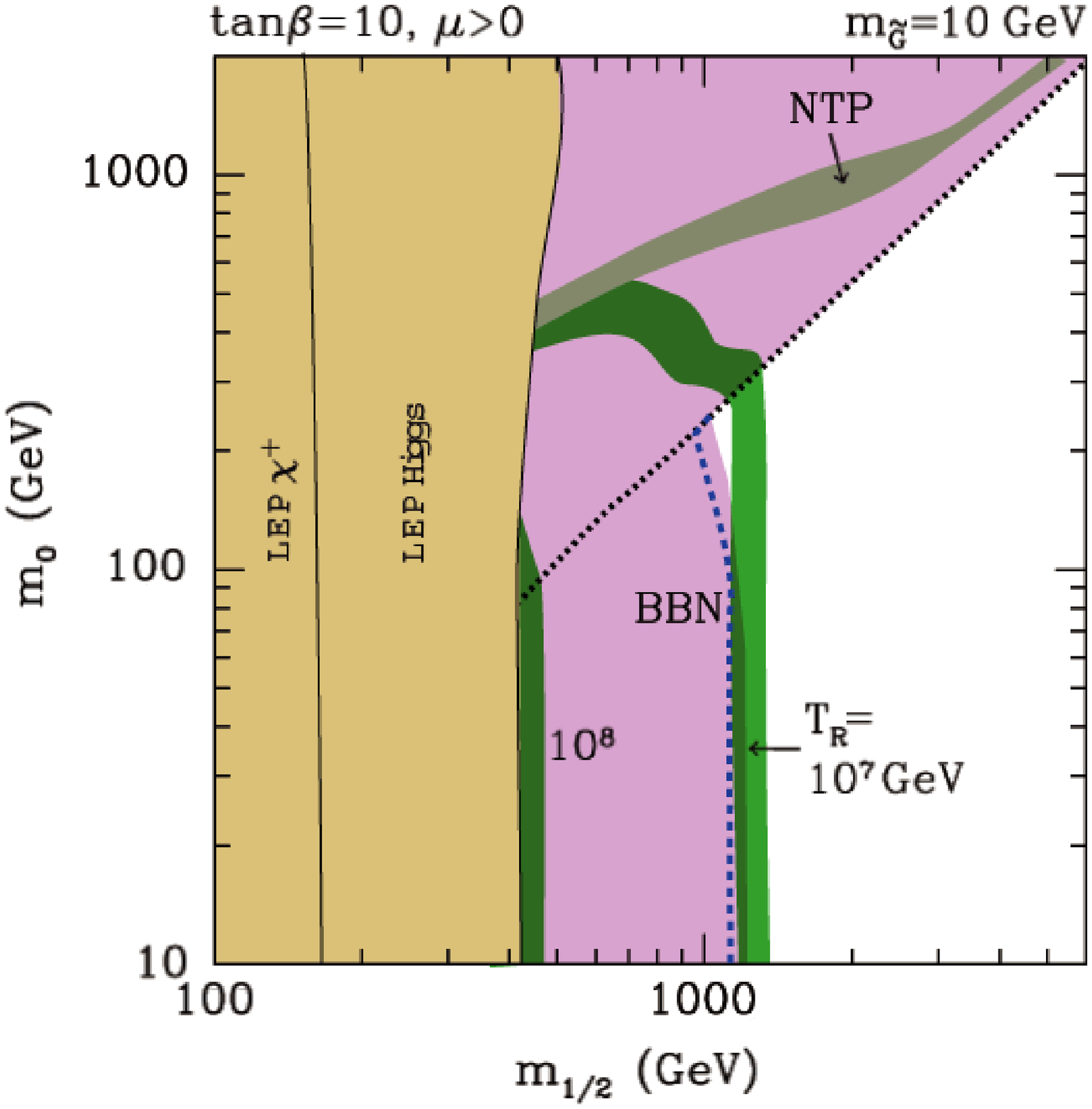}
    & 
   \includegraphics[width=0.5\textwidth]{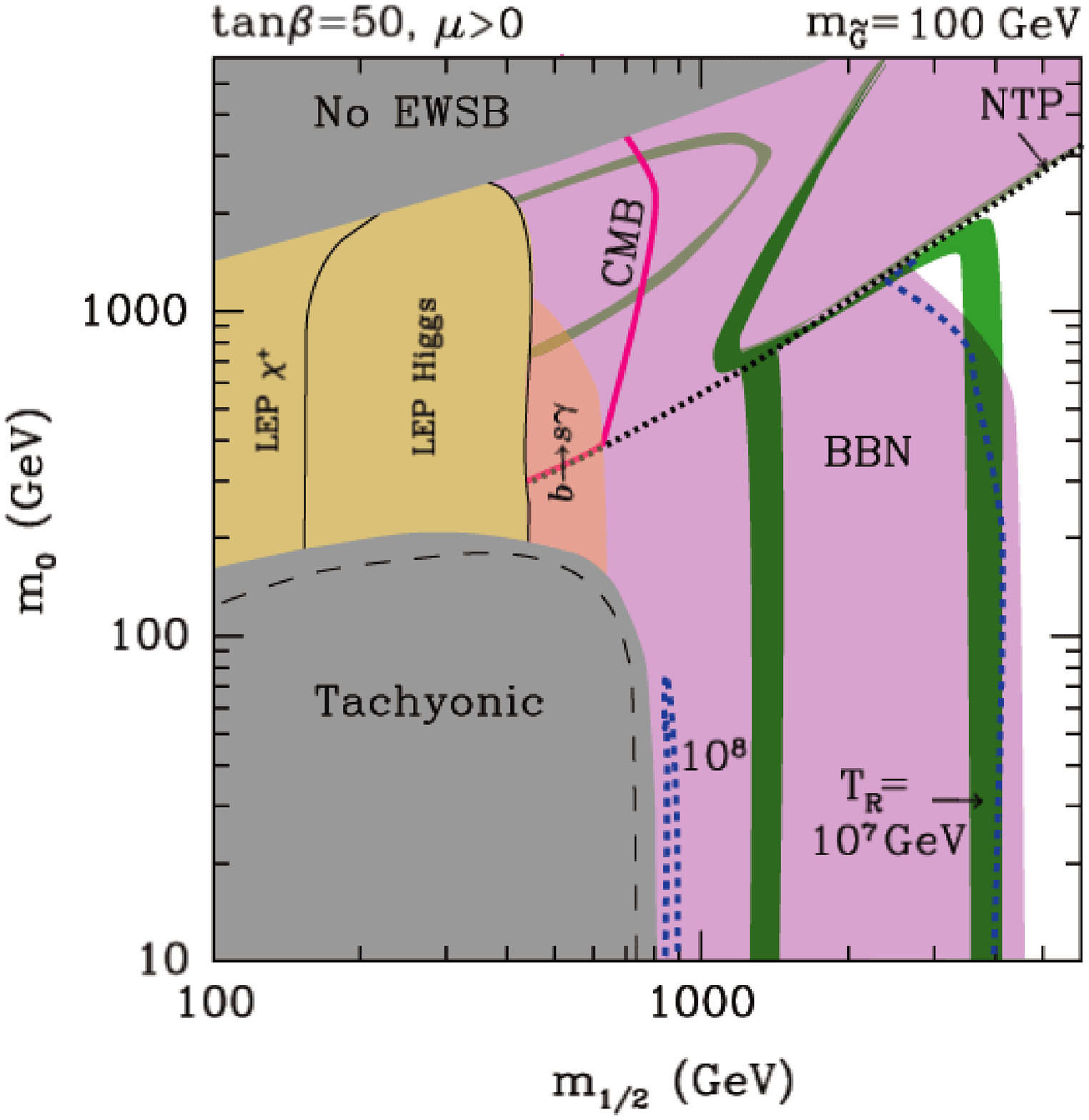}
  \end{tabular}
  \end{center}
  \caption{The same as fig.~\protect\ref{fig:bbn} but for a fixed
    gravitino mass, 
     $\tanb=10$, $\mgravitino=10\gev$ (left window) and
    $\tanb=50$, $\mgravitino=100\gev$ (right window). }
  \label{fig:fixedmG}
\end{figure}

As regards the total gravitino relic abundance $\abundg$, we apply the 
$3\,\sigma$ range derived from WMAP 5 year data~\cite{Komatsu:2008hk}
\dis{
0.091< \abundg < 0.128,
\label{eq:cdmwmap}
}
which in the figures below will be marked as green bands and labeled ``$\abundg$''. 

As previously in~\cite{Cerdeno:2005eu}, we also include the bound on
the possible distortion in the nearly perfect black-body shape of the
CMB spectrum~\cite{hu93} by the injection of energetic photons into
the plasma.  However we note that this constraint (delineated with
magenta line with a label ``CMB'' over it) seems generally less
important than that due to the BBN~\cite{Sigl:1995kk, Lamon:2005jc}.

\subsection{Numerical results}

In figs.~\ref{fig:bbn} and~\ref{fig:fixedmG} we give the updated
results with the constraints and the relic abundance of the gravitino
in the same format as used in~\cite{Cerdeno:2005eu} to facilitate the
comparison. Generally, the regions in white are left allowed after applying collider
and BBN constraints. Green bands denote ranges of parameters where the
total gravitino abundance reproduces the DM
abundance~(\ref{eq:cdmwmap}), while in the light green bands (marked
``NTP'') only the NTP part of the gravitino abundance agrees with that range.

In fig.~\ref{fig:bbn} we present two cases with gravitino mass
related to the soft scalar mass $\mzero$.
In the left window we take $\tanb=10$ and $\mgravitino=\mzero$ and 
in the right window $\tanb=50$ and $\mgravitino=0.2\mzero$.
In fig.~\ref{fig:fixedmG} we present two cases with a fixed
$\mgravitino$. In the left window we fix $\tanb=10$ and $\mgravitino=10\gev$
while in the right one $\tanb=50$ and $\mgravitino=100\gev$.

We can see that, as before, the whole neutralino NLSP region (above
the black dotted diagonal line) is again ruled out, as well as a part
of the stau NLSP region corresponding to smaller $\mhalf$ (and
therefore smaller $\mstau$ and hence larger stau lifetimes). By
comparing with the corresponding figures in~\cite{Cerdeno:2005eu}, we
can see that including the bound state effects causes the allowed
region of stau NLSP consistent with DM abundance (green band) to be
more constrained, and also strengthens an upper bound on $\treh$ to
%%%
\dis{\treh<{\rm a~few~}\times10^7 \gev.
\label{trehupperbound}
}
%%%
(For comparison, without the bound state effect, the bound was a few
times $10^8 \gev$~\cite{Cerdeno:2005eu}.) This bound is consistent
with the result of ref.~\cite{Pradler:2006hh,Pradler:2007is} where the
bound state effect was mimicked by making a simple and approximate
constraint on the lifetime of the stau, $\tau<5\times
10^3\sec$. However, as we shall now show, in a number of cases the
stau lifetime that can be consistent with avoiding strong bound-state
catalytic effect can be almost an order of magnitude larger.
Furthermore, by applying more conservative bounds on 
${^6}{\! \rm Li}/{^7}{\!\rm Li}$, the upper
bound~(\ref{trehupperbound}) can in some cases be violated by up to an
order of magnitude.

\begin{figure}[!t]
  \begin{center}
  \begin{tabular}{c c}
   \includegraphics[width=0.5\textwidth]{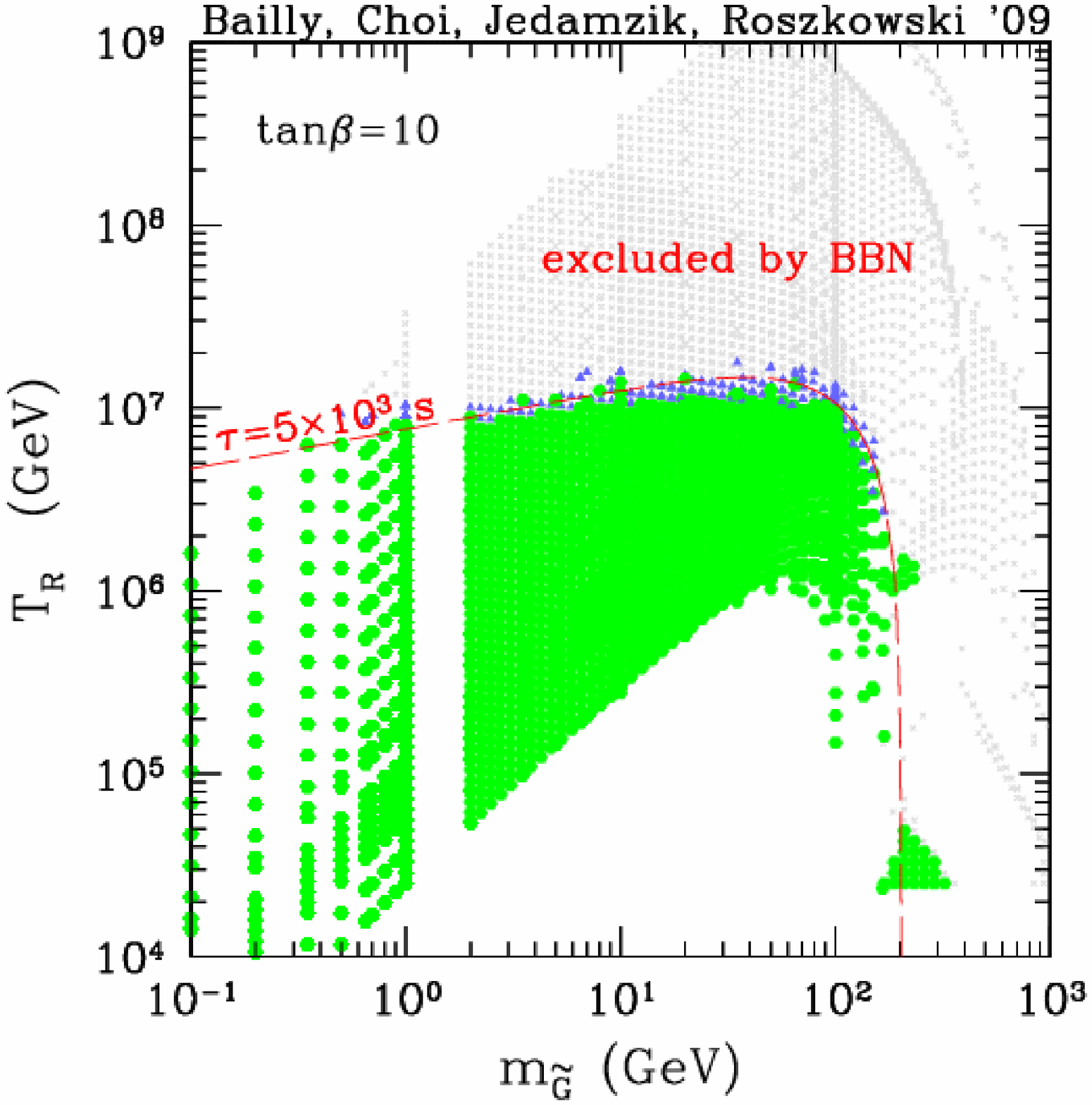}
    & 
   \includegraphics[width=0.5\textwidth]{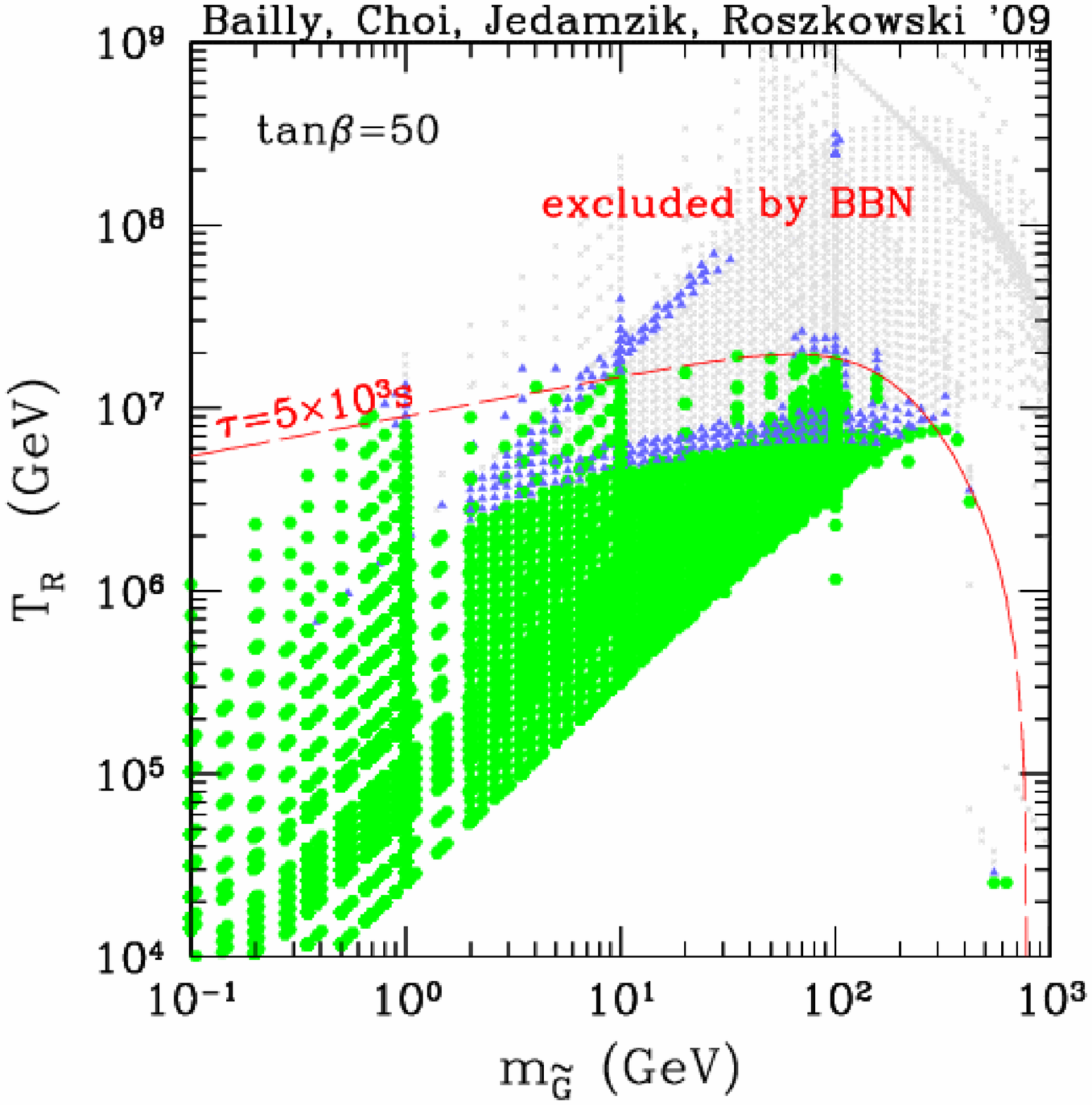}
 \end{tabular}
  \end{center}
  \caption{\small The $\treh$ vs. $\mgravitino$ plane for $\tanb=10$
(left) and $\tanb=50$ (right). Data points which survived all BBN and
experimental constraints are shown with green dots, while grey dots
are disallowed by the BBN while allowed by collider constraints.  Blue
points are added to the green when we use a more conservative limit 
$^6{\rm Li}/^7{\rm Li} < 0.66$.  $\treh$ is determined so that the
total (thermal and non-thermal) production of gravitinos satisfies the
WMAP value; here we used $\abundg =\abundgtp+\abundgntp=0.11$.  We
also show the lifetime contour of stau, $\tau=5\times 10^3\sec$, which
was obtained using the specific relation between stau mass and gaugino
masses as well as the approximate relic density of the $\stau$ in the
 CMSSM.}
  \label{fig:mGTR_stau}
\end{figure}

In order to examine the upper bound on $\treh$ more closely, in
fig.~\ref{fig:mGTR_stau} we plot $\treh$ vs. $\mgravitino$ for all the
points in our scans which satisfy the constraints from BBN and
collider experiment and for $\tanb=10$ (left) and $\tanb=50$ (right).
In these plots we use several choices of the gravitino mass: linear in
the scalar mass ($0.2\mzero, 0.4\mzero$), linear in the gaugino mass
($0.2\mhalf, 0.4\mhalf$) or constant (24 values between $0.1 \gev$ and
$100\gev$). For each point, $\treh$ is calculated using
eqs.~(\ref{abundgtp:eq}) and~(\ref{eq:oh2relation}) for a given
gravitino mass and for the mass spectra and the yield of NLSP at each
point of parameter space, imposing $\abundg=0.11$.  Points marked by
green dots survive all the constraints from experiments and cosmology
while those marked by grey dots are allowed by experiments but
disallowed by BBN constraints from light element abundances. Again, we can see that, for both
choices of $\tanb$, we find the upper bound~(\ref{trehupperbound}).
This result does not change if one alters collider constraints since
the bound is set by the $^6{\rm Li}$ abundance.
\begin{figure}[!t]
  \begin{center}
  \begin{tabular}{c c}
   \includegraphics[width=0.5\textwidth]{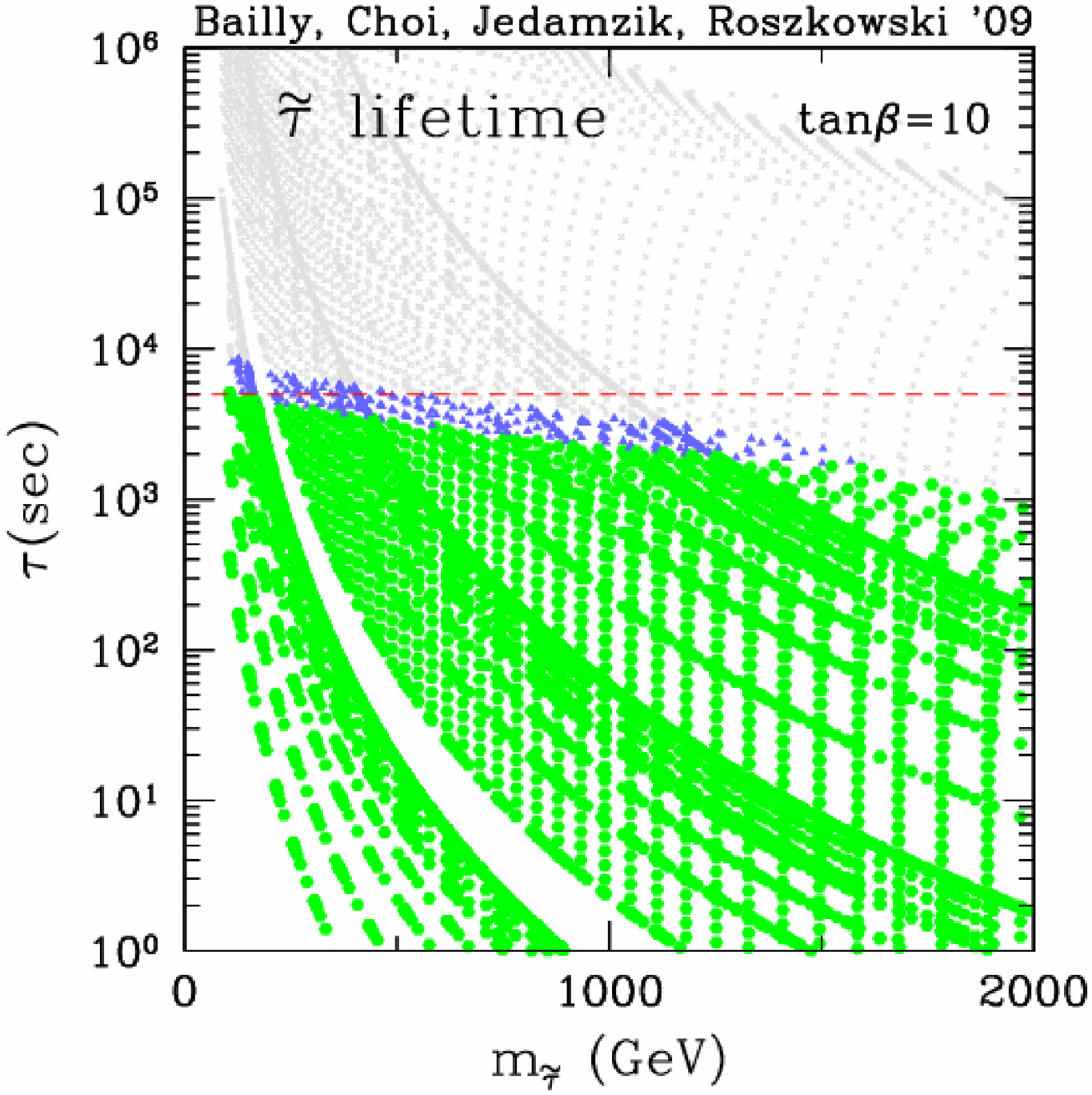}
    & 
   \includegraphics[width=0.5\textwidth]{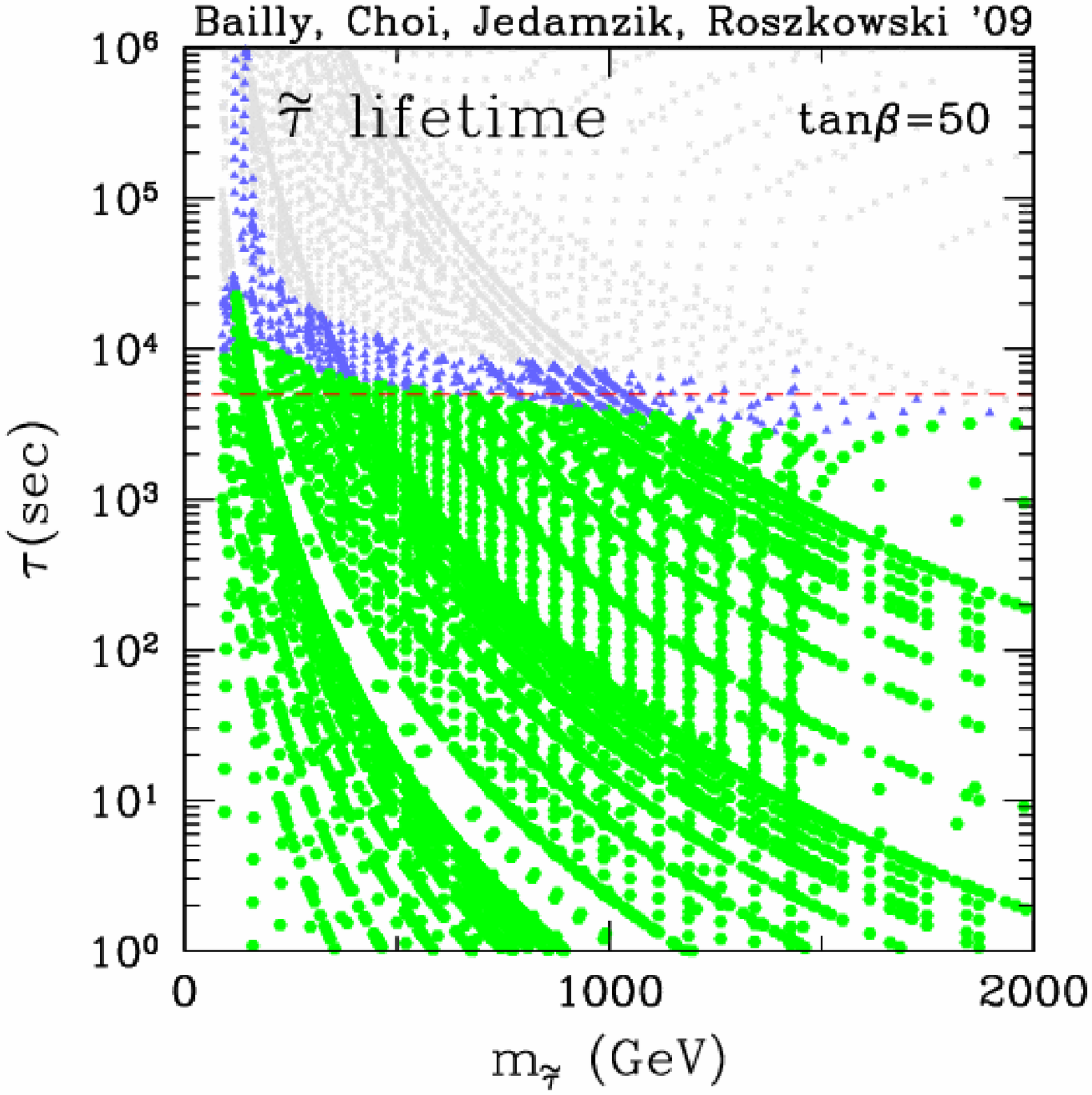}
  \end{tabular}
  \end{center}
  \caption{\small The same as fig.~\ref{fig:mGTR_stau} but in the
  plane of lifetime and mass of stau for $\tanb=10$ (left) and  
  $\tanb=50$ (right).} 
  \label{fig:lifetime_mstau}
\end{figure}

However, when we use the conservative bound on $^6{\rm Li}/^7 {\rm
Li}< 0.66$ then we find new isolated, albeit rather small allowed
regions (marked by blue dots), where the reheating temperature can go
up to $3\times 10^8\gev$. (They correspond to vertical blue dashed
region around $\mhalf\sim 1 \tev$ and small $\mzero$ in the right
window in figs.~\ref{fig:bbn} and~\ref{fig:fixedmG}). Clearly, those
points evade the simple bound $\tau<5\times 10^3\sec$ (marked with a
red dashed line).  On the other hand, this region corresponds to a
rather low stau mass around $100\gev$ (but still above the LEP limit).
This can be seen in fig.~\ref{fig:lifetime_mstau} where we plot stau
lifetime for the data points shown in fig.~\ref{fig:mGTR_stau}.  We
can see that the (green) points allowed by $^6{\rm Li}/^7{\rm Li}<0.1$
are located below the stau lifetime of $5 \times10^3 \sec$ for
$\tanb=10$ and $3\times10^4 \sec$ for $\tanb=50$. It is also clear
that, at smaller $\tanb$ (left panel) and larger $\mstau$ the bound
$\tau<5\times 10^3\sec$ can actually be too weak.  On the other hand,
assuming the conservative bound $^6{\rm Li}/^7{\rm Li}<0.66$ (blue
points), at stau mass around $100\gev$, for $\tanb=50$ the lifetime
can exceed $10^4 \sec$ even by a few orders of magnitude. Such low
stau mass will be accessible to the LHC, thus allowing one to
scrutinize such cases fairly easily.

\begin{figure}[!t]
  \begin{center}
  \begin{tabular}{c c}
   \includegraphics[width=0.5\textwidth]{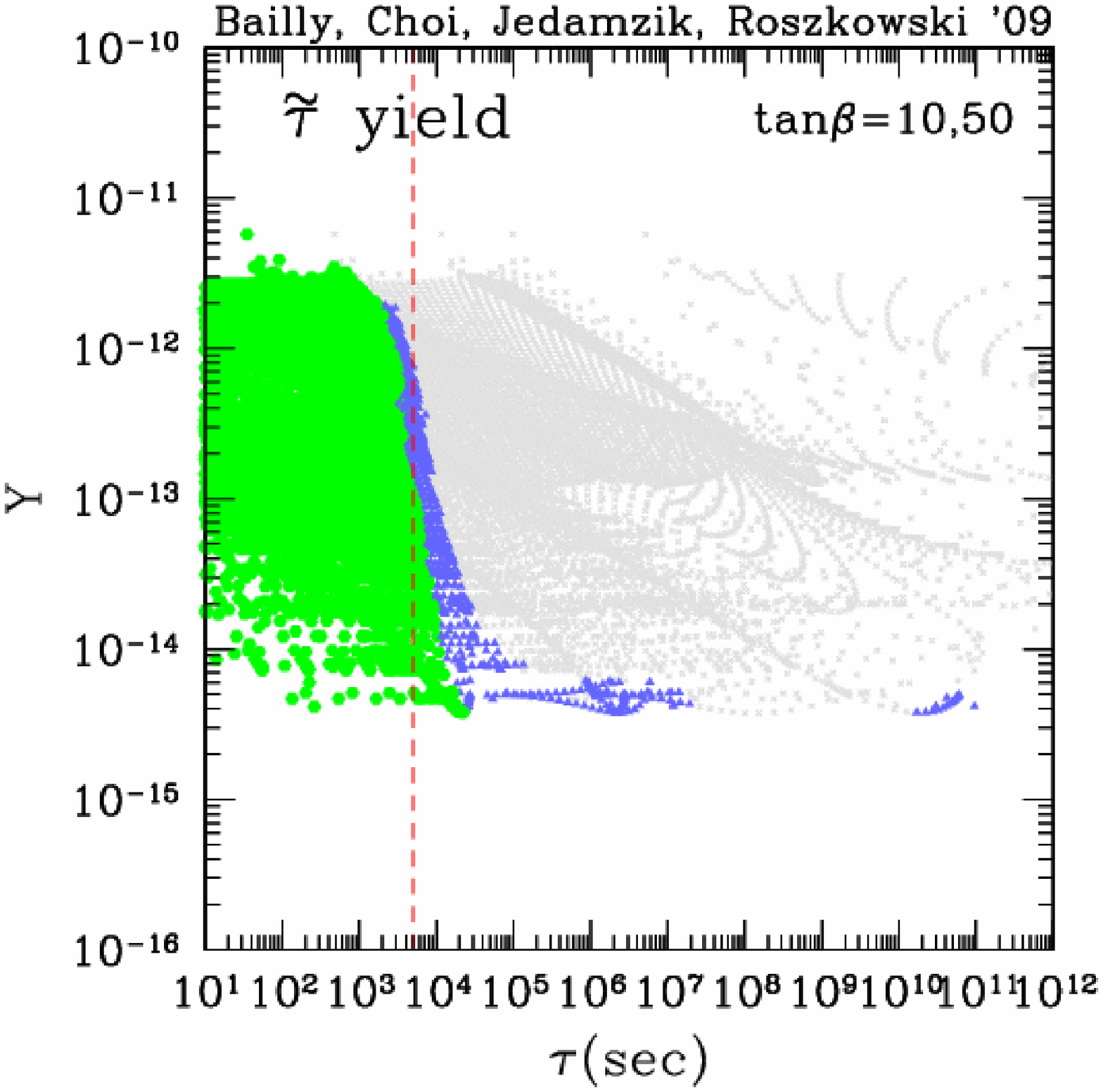}
    & 
   \includegraphics[width=0.5\textwidth]{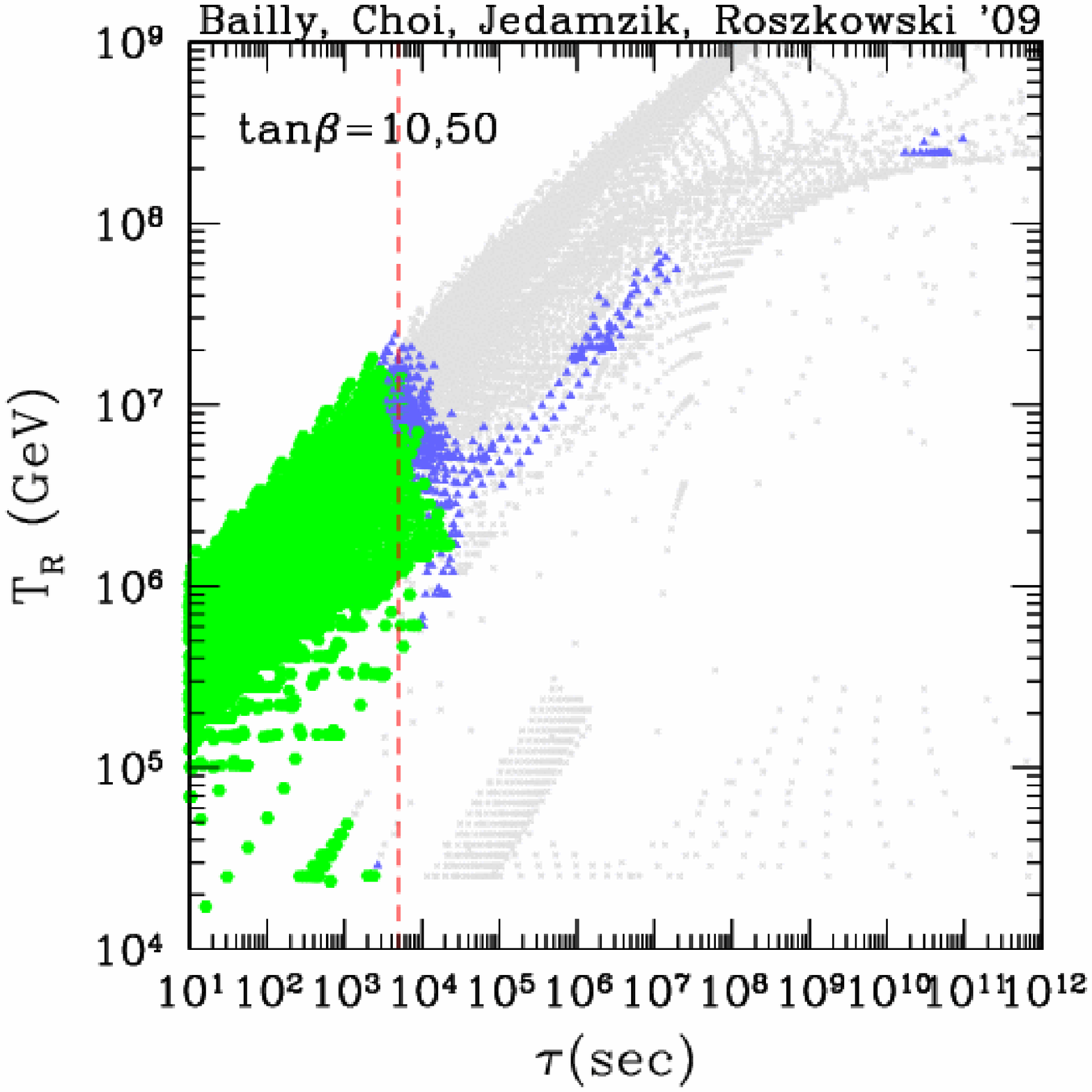}
  \end{tabular}
  \end{center}
\caption{\small The same as fig.~\ref{fig:mGTR_stau} but in the plane
   of the stau yield $Y\equiv n/s$ (left window) and of $\treh$ (right
   window) vs. stau lifetime for
  both $\tanb=10$ and $\tanb=50$. 
}
\label{relicY}
\end{figure}

The reason why the points corresponding to stau NLSP with such a long
lifetime are allowed is that, as the stau mass decreases, its yield
$Y$ also decreases and can drop below $10^{-14}$.  In this case the
BBN constraints from the bound state effects can be avoided even with
stau lifetime as long as $3\times 10^4\sec$ (green points).

To see this, we present the left window of fig.~\ref{relicY}. 
Assuming a standard (more conservative) limit
on ${^6}{\! \rm Li}/{^7}{\!\rm Li}$ we find green (blue) points
corresponding to the lifetime of up to $3\times 10^4 \sec$ ($2\times
10^7 \sec$) and the stau yield of less than $10^{-14}$.  For lifetimes
longer than $10^7 \sec$ the constraint due to violating the upper
limit on $^3{\rm He}/{\rm D}$ due to $^4{\rm He}$ photo-disintegration
becomes severe~\cite{Jedamzik:2007qk}.  On the other hand, for much
longer lifetimes, between $10^{10-11}\sec $, we again find some
isolated allowed regions of (blue) points. They become allowed because
the mass difference between the gravitino LSP and the stau NLSP is
very small there and the electromagnetic energy released is not enough
to be a potential problem for BBN.

Finally, in the right window of fig.~\ref{relicY} we show $\treh$
vs. stau lifetime for all the cases we have considered, and which
summarizes many of our points made above. In particular, with the
ranges of primordial light element abundances adopted in this analysis
(green points), the upper limit on $\treh$ scales with the stau
lifetime $\tau$ but does not exceed the
limit~(\ref{trehupperbound}). Allowing the more conservative limit
on ${^6}{\! \rm Li}/{^7}{\!\rm Li}$ pushes it up to nearly $10^8\gev$,
with some isolated points tolerating $\treh$ as high as $3\times10^8\gev$, as stated above.
   
%=====================
\section{Summary}
%\paragraph{Summary.}
%=====================
\label{sect:summary}

We have re-analyzed the gravitino as dark matter in the Universe
in the framework of the CMSSM, taking into account of a number of recent
improvements in the calculation. The improvements concern the 
thermal production of
gravitinos, including the decay allowed by thermal masses as well as
scatterings, the computation of the hadronic spectrum from NLSP decay
including the correct implementation of left-right stau mixing, as well as an
updated BBN calculation fully treating the effects of bound state 
effects between charged
massive particles and nuclei.  
We found that the over-abundance of $^6{\rm Li}$
from bound state effects puts more severe constraints on allowed ranges
of CMSSM parameters than before and puts an upper bound on $\treh$ 
at a few times $10^7 \gev$ and the stau lifetime
less than around $3\times 10^4 \sec$.

We also analyzed the impact of applying a more conservative upper
limit on $^6{\rm Li}/^7{\rm Li}$.  In this case a new region at small
stau mass opens up because of the reduced relic abundance of the stau,
even though the lifetime of the stau is much longer than $10^4
\sec$, and can reach up to $5\times 10^7 \sec$, in some cases even between
$10^{10}$ and $10^{11}\sec$. In these latter 
cases, which are easily testable at the LHC, 
$\treh$ can exceed $10^8 \gev$. 

\acknowledgments K.-Y.C. is supported by the Ministerio de Educacion y
 Ciencia of Spain under Proyecto Nacional FPA2006-05423 and by the
 Comunidad de Madrid under Proyecto HEPHACOS, Ayudas de I+D
 S-0505/ESP-0346. K.-Y.C. and L.R. would like to thank the European
 Network of Theoretical Astroparticle Physics ILIAS/ENTApP under
 contract number RII3-CT-2004-506222 for financial support.

\end{document}